\documentclass[traditabstract,longauth]{aa}
\usepackage{natbib}
\usepackage{epsfig}
\usepackage{txfonts}
\newcommand{\corot}{\emph{CoRoT}}

\newcommand{\Ctthree}{\emph{CoRoT-3b}}

\newcommand{\Ctsix}{\emph{CoRoT-6b}}

\newcommand{\Ctfourt}{\emph{CoRoT-14b}}

\newcommand{\Cttw}{\emph{CoRoT-20b}}
\newcommand{\Cttws}{\emph{CoRoT-20}}
\newcommand{\SRa}{\emph{SRa03}}

\newcommand{\sophie}{\emph{SOPHIE}}
\newcommand{\harps}{\emph{HARPS}}

\newcommand{\fies}{\emph{FIES}}

\newcommand{\mass}{\emph{2-MASS}}

\def\ms{\,m\,s$^{-1}$}         
\def\kms{\hbox{\,km\,s$^{-1}$}}       
\def\gcm{\hbox{\,g\,cm$^{-3}$}}       
\def\vsini{\hbox{$v$\,sin\,$i$}}      
\def\halpha{\hbox{$H\alpha$}}      
\def\Ms{\hbox{$M_{\star}$}}             
\def\Rsun{\hbox{$R_{\odot}$}}
\def\Rs{\hbox{$R_{\star}$}}

\def \Rp{$R_{\rm p}$}
\newcommand{\vrad}{$v_{\rm rad}$} 
\newcommand{\teff}{T$_{\rm eff}$}
\newcommand{\logg}{log {\it g}}
\newcommand{\vmic}{v$_{mic}$}
\newcommand{\vmac}{v$_{macro}$}

\newcommand{\feh}{[Fe/H]}
\newcommand{\mtier}{$M_\star^{1/3}/R_\star$} 
\newcommand{\MJ}{{\it M}$_{\rm Jup}$}
\newcommand{\ME}{{\it M}$_\oplus$}
\newcommand{\RJ}{{\it R}$_{\rm Jup}$}

\newcommand{\Msun}{$M_{\odot}$}             

\begin{document}
\title{Transiting exoplanets from the CoRoT space mission\thanks{The CoRoT space mission, launched on December 27th 2006, has been developed and is operated by CNES, with the contribution of Austria, Belgium, Brazil , ESA (RSSD and Science Programme), Germany and Spain.}}
\subtitle{XX. CoRoT-20b: A very high density, high eccentricity transiting giant planet }

\author{
M. Deleuil\inst{\ref{inst1}}
\and A.S. Bonomo\inst{\ref{inst1}}
\and S. Ferraz-Mello \inst{\ref{inst10}}
\and A. Erikson \inst{\ref{inst9}}
\and F. Bouchy\inst{\ref{inst4},\ref{inst5}}
\and M. Havel\inst{\ref{inst19}}
\and S. Aigrain \inst{6}
\and J.-M. Almenara \inst{\ref{inst1}} 
\and R. Alonso\inst{17} 
\and M. Auvergne\inst{\ref{inst2}}
\and A. Baglin\inst{\ref{inst2}} 
\and P. Barge \inst{\ref{inst1}} 
\and P. Bord\'e\inst{\ref{inst3}} 
\and H. Bruntt\inst{\ref{inst21}} 
\and J. Cabrera\inst{9} 
\and Sz. Csizmadia\inst{\ref{inst9}}
\and C. Damiani \inst{\ref{inst1}}
\and H.J., Deeg\inst{\ref{inst7},\ref{inst7b}}
\and R. Dvorak\inst{\ref{inst8}} 
\and M. Fridlund\inst{\ref{inst11}}
\and G. H\'ebrard\inst{\ref{inst4},\ref{inst5}}
\and D. Gandolfi\inst{\ref{inst11}}
\and M. Gillon\inst{\ref{inst12}} 
\and E. Guenther\inst{\ref{inst13}}
\and T. Guillot\inst{\ref{inst19}}
\and A. Hatzes\inst{\ref{inst13}}
\and L. Jorda\inst{\ref{inst1}}
\and A. L\'eger\inst{\ref{inst3}}
\and H. Lammer\inst{\ref{inst14}}
\and T. Mazeh\inst{\ref{inst16}}
\and C. Moutou, C.\inst{\ref{inst1}} 
\and M. Ollivier\inst{\ref{inst3}}
\and A. Ofir\inst{\ref{inst22}}
\and H. Parviainen\inst{\ref{inst7},\ref{inst7b}}
\and D. Queloz\inst{\ref{inst17}}
\and H. Rauer\inst{\ref{inst9}}
\and A. Rodr\'{\i}guez\inst{\ref{inst10}} 
\and D. Rouan\inst{\ref{inst2}}
\and A. Santerne\inst{\ref{inst1}}
\and J. Schneider\inst{\ref{inst20}}
\and L. Tal-Or\inst{\ref{inst16}}
\and B. Tingley\inst{\ref{inst7},\ref{inst7b}}
\and J. Weingrill\inst{\ref{inst18}}
\and G. Wuchterl\inst{\ref{inst13}}
}

\institute{
Laboratoire d'Astrophysique de Marseille, 38 rue Fr\'ed\'eric Joliot-Curie, 13388 Marseille cedex 13, France\label{inst1}
\and LESIA, Obs de Paris, Place J. Janssen, 92195 Meudon cedex, France\label{inst2}
\and Institut d'Astrophysique Spatiale, Universit\'e Paris XI, F-91405 Orsay, France\label{inst3}
\and Observatoire de Haute Provence, 04670 Saint Michel l'Observatoire, France\label{inst4}
\and Institut d'Astrophysique de Paris, 98bis boulevard Arago, 75014 Paris, France\label{inst5}
\and Department of Physics, Denys Wilkinson Building Keble Road, Oxford, OX1 3RH\label{inst6}
\and Instituto de Astrofisica de Canarias, E-38205 La Laguna, Tenerife, Spain\label{inst7} 
\and Universidad de La Laguna, Dept. de Astrof\'\i sica, E-38200 La Laguna, Tenerife, Spain\label{inst7b}
\and University of Vienna, Institute of Astronomy, T\"urkenschanzstr. 17, A-1180 Vienna, Austria\label{inst8}
\and Institute of Planetary Research, German Aerospace Center, Rutherfordstrasse 2, 12489 Berlin, Germany\label{inst9}
\and IAG, Universidade de Sao Paulo, Brazil \label{inst10}
\and Research and ScientiÞc Support Department, ESTEC/ESA, PO Box 299, 2200 AG Noordwijk, The Netherlands\label{inst11} 
\and University of Li\`ege, All\'ee du 6 ao\^ut 17, Sart Tilman, Li\`ege 1, Belgium\label{inst12}
\and Th\"uringer Landessternwarte, Sternwarte 5, Tautenburg 5, D-07778 Tautenburg, Germany\label{inst13}
\and Space Research Institute, Austrian Academy of Science, Schmiedlstr. 6, A-8042 Graz, Austria\label{inst14} 
\and School of Physics and Astronomy, Raymond and Beverly Sackler Faculty of Exact Sciences, Tel Aviv University, Tel Aviv, Israel\label{inst16}  
\and Observatoire de l'Universit\'e de Gen\`eve, 51 chemin des Maillettes, 1290 Sauverny, Switzerland\label{inst17} 
\and Observatoire de la C\^ote dÕ Azur, Laboratoire Cassiop\'ee, BP 4229, 06304 Nice Cedex 4, France\label{inst19}
\and LUTH, Observatoire de Paris, CNRS, Universit\'e Paris Diderot; 5 place Jules Janssen, 92195 Meudon, France \label{inst20}
\and Department of Physics and Astronomy, Aarhus University, 8000 Aarhus C, Denmark\label{inst21}
\and Wise Observatory, Tel Aviv University, Tel Aviv 69978, Israel\label{inst22}
}
\date{Received ; accepted }

\abstract{We report the discovery by the \corot\ space mission of a new giant planet, \Cttw. The planet has a mass of 4.24 $\pm$ 0.23 \MJ\ and a radius of 0.84 $\pm$ 0.04 \RJ. With a mean density of 8.87 $\pm$ 1.10 \gcm, it is among the most compact planets known so far.  Evolution models for the planet suggest a mass of heavy elements of the order of  800 $M_{\oplus}$ if embedded in a central core, requiring a revision either of the planet formation models or of planet evolution and structure models.  We note however that smaller amounts of heavy elements are expected from more realistic models in which they are mixed throughout the envelope. The planet orbits a G-type star with an orbital period of 9.24 days and an eccentricity of 0.56. The star's projected rotational velocity is \vsini\ = 4.5 $\pm$ 1.0 \kms, corresponding to a spin period of 11.5 $\pm$ 3.1 days if its axis of rotation is perpendicular to the orbital plane. In the framework of Darwinian theories and neglecting stellar magnetic breaking, we calculate the tidal evolution of the system and show that \Cttw\ is presently one of the very few Darwin-stable planets that is evolving towards a triple synchronous state with equality of the orbital, planetary and stellar spin periods.
}

\titlerunning{CoRoT-20b: A very high density, high eccentricity transiting planet}
\authorrunning{M. Deleuil}

\maketitle

\section{Introduction}
The existence of a planet population at very short orbital distance, $a < 0.1 AU$ typically, with its wide range of orbital and physical properties is an intriguing phenomenon. In-situ formation of such massive bodies so close to their host-star at a location where the circumstellar material is depleted and warm, appears indeed highly unlikely. Planet migration from further away distances where solid material is abundant, triggered by gravitational interactions, is invoked to account for this population. The exact process is still under debate but two major mechanisms are put forward: gradual planet migration due to interaction with the circumstellar gas disk  \citep{Lin96, Papaloizou07} or planet-planet or planet-companion star interactions combined with tidal dissipation \citep[][e.g.]{Rasio96, Fabrycky07, Nagasawa08}. Whatever the exact nature of the formation path, these planets have undergone a significant orbital evolution since the time of their formation. Their current properties provide valuable hints helping to better understand their orbital evolution, especially the planet's orbit eccentricity and spin-orbit alignment of the system. 

Up to now, transit surveys have been more sensitive to planets at very short orbital period. The situation has recently started to change thanks to the extended temporal coverage of ground-based transit surveys and the advent of space missions, \corot\ \citep{Baglin09,Deleuil11} and Kepler \citep{Borucki10}. As a consequence, the number of planets with orbital periods greater than a few days has significantly increased over the past two years. While the mean eccentricity for the close-in planets is close to zero, the transiting giant planets at larger orbital distance display a much wider range in eccentricity, a picture more consistent with the sample of planets found by radial velocity surveys. These trends 
favor a third body induced migration with tidal circularization of an initial eccentric and possibly high-obliquity orbit \citep[][e.g.]{Winn10,Matsumura10,Pont11}.  A consequence of this orbital evolution is the tidal destruction of the planet which spirals down onto the star in the life time of the system \citep{Gonzalez97,Jackson09}, a dramatic destruction that appears being the fate of the vast majority of the transiting planets \citep{Matsumura10}.

In this paper, we report on the discovery of \Cttw, a new member of the hot-Jupiter class population. The planet transits its G-type parent star every 9.24 days, along an orbit with a high eccentricity. The \corot\ observations are 
presented in Section~\ref{LCurve}. The accompanying follow-up observations and their results are described in Section~\ref{FUp} and Section~\ref{StellarP} for the host-star analysis. The final system parameters are derived in Section~\ref{SysParam}. We then discuss the properties of this new planet in Section~\ref{SysProp}. We investigate its orbital evolution and fate but also its internal structure that rises new questions on the nature of such compact object. 
\begin{figure*}[ht]
\begin{center} 
\includegraphics[width=17cm,height=4cm]{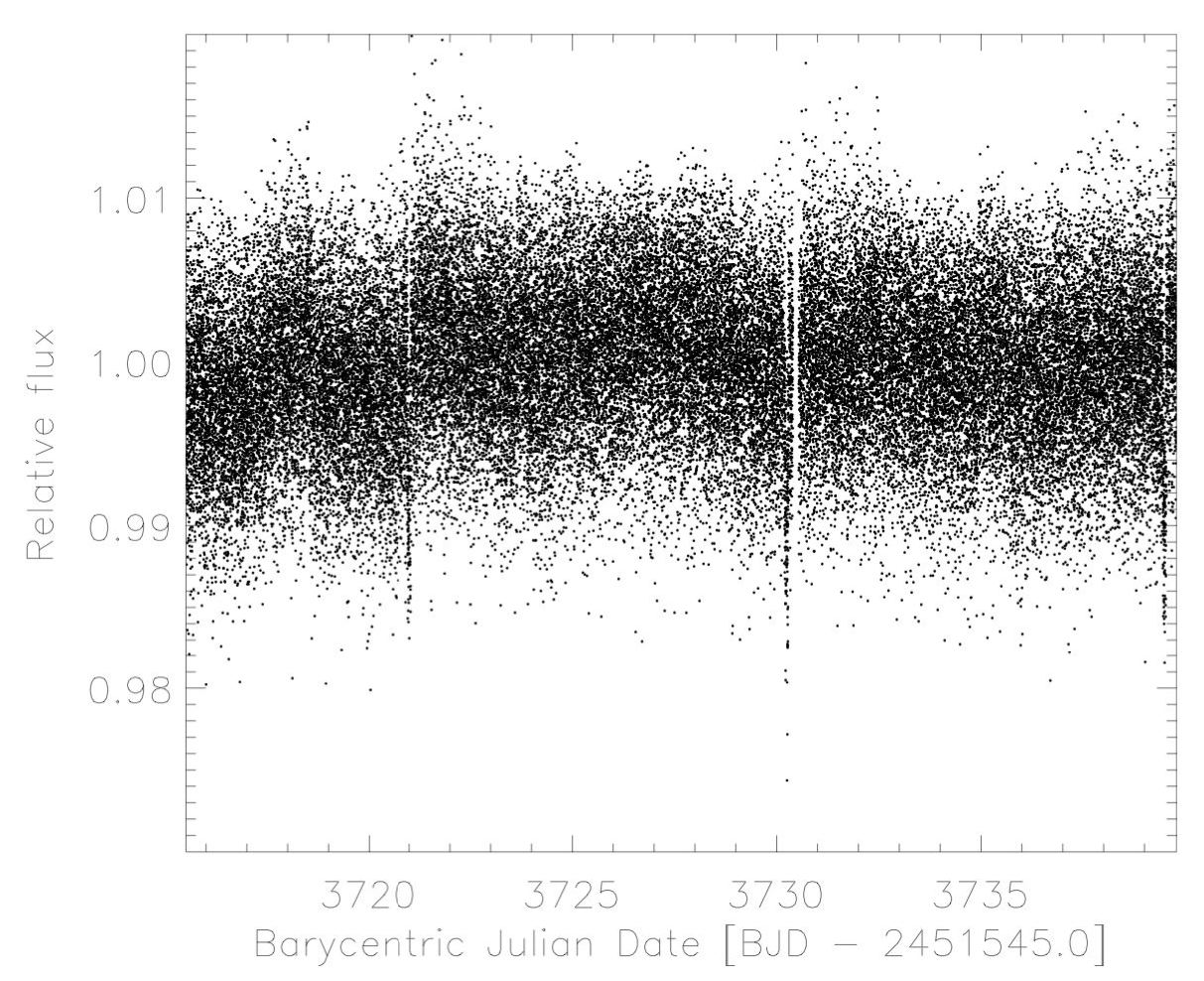}
\caption{The 24.278-days long \Cttws\ reduced light curve at a constant 512-sec time sampling.}
\end{center}
\label{LCTotal}
\end{figure*}
 
\section{\Cttw\ Light curve}
\label{LCurve}
The planet has been discovered in one of the fields observed by the \corot\ satellite \citep{Baglin09,Deleuil11} in the so-called {\sl anti-center} direction. This field, labeled as \SRa, was monitored for 24.278 days starting on 1 
March 2010. 
As a consequence of the DPU1 break down that took place on March 2009, the number of targets actually photometrically monitored by the instrument was reduced to a maximum of 6000 stars only. The released telemetry is further used to oversample a much larger number of targets than it was initially possible, up to magnitude $\simeq$ 15. While not among the brightest stars of the field (Table~\ref{startable}), \Cttws\ benefited this new opportunity and its observation was performed with the regular 32~sec time sampling. It was also bright enough to allow for three-color photometry. Two transits were detected in its light curve
by the {\sl Alarm mode} pipeline \citep{Surace08}. The target was flagged as a good planetary candidate and put among the highest priorities for follow-up observations. 

\begin{table}[h]
\caption{ \Cttw\ IDs, coordinates and magnitudes.}            
\centering        
\begin{minipage}[!]{7.0cm}  
\renewcommand{\footnoterule}{}     
\begin{tabular}{lcc}       
\hline\hline                 
CoRoT window ID & SRa03\_E2\_0999 \\
CoRoT ID & 315239728 \\
USNO-B1 ID  & 0902-0091920 \\
2MASS ID   &  06305289+0013369 \\
GSC2.3 ID &  \\
\\
\multicolumn{2}{l}{Coordinates} \\
\hline            
RA (J2000)  & 97.720434 \\
Dec (J2000) &  0.22692\\
\\
\multicolumn{3}{l}{Magnitudes} \\
\hline
\centering
Filter & Mag & Error \\
\hline
B$^a$  & 15.31 & \\
V$^a$  & 14.66 & \\
J$^b$  & 12.991 & 0.023 \\
H$^b$  & 12.652 & 0.026 \\
K$^b$  &  12.512  &  0.027\\
\hline\hline
\vspace{-0.5cm}
\footnotetext[1]{from USNO-B1 - Provided by Exo-Dat \citep{Deleuil09};}
\footnotetext[2]{from \mass\ catalog.}
\end{tabular}
\end{minipage}
\label{startable}      
\end{table}

The light curve of \Cttw\ is displayed in Fig~\ref{LCTotal}. It shows a star rather quiet photometrically speaking, with no special prominent feature. Three transits are clearly visible with a period of 9.24 days and a depth slightly shallower than 1\%.  For the detailed analysis, we used the light curve reduced by the \corot\ calibration pipeline. It corrects for the main instrumental effects such as the CCD zero offset and gain, the background light and the spacecraft jitter \citep[see][]{Auvergne09}. Portions of the light curve that were flagged by the pipeline as affected by particle impacts during the South Atlantic Anomaly crossing, were removed and ignored in the analysis. In total, the light curve consists of 56869 photometric measurements. It gives a corresponding duty cycle of ~88\%.
\section{Follow-up observations}
\label{FUp}
\begin{figure}[h]
\begin{center} 
\includegraphics[height=4cm,width=5cm]{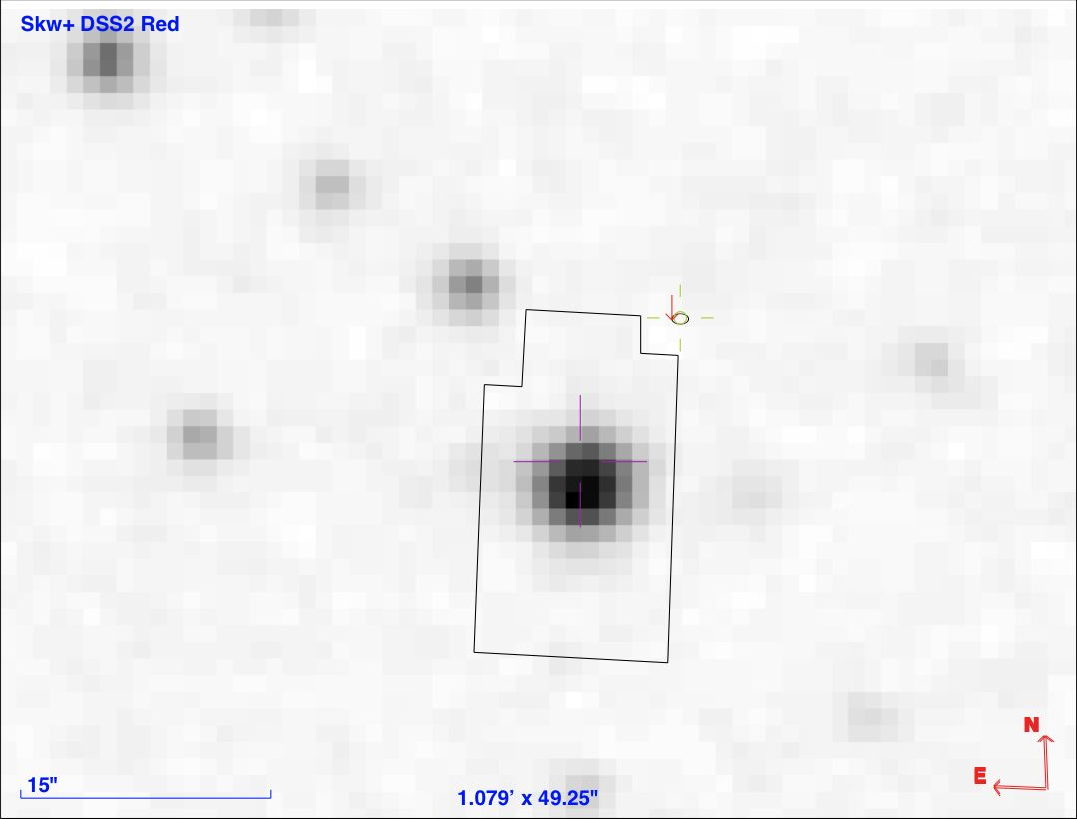}
\end{center}
\caption{Image of the DSS showing  \Cttws\ and its environment. The photometric mask used for \corot\ observations is overplotted on the target.}
\label{Skyview}
\end{figure}

A photometric time-series of the star was obtained at the {\emph Wise} observatory on November 14, 2010 in order to check whether an unknown nearby eclipsing binary could be the source of the transits \citep{Deeg09}. The detection of a transit ingress excluded this configuration at the spatial resolution of {\emph Wise}. The observed time of the ingress, with first contact at 2455515.510$\pm$ 0.007 HJD was then used to refine the period of \Cttw, towards the value quoted in Table~\ref{AllParams}. Ground-based images from both {\emph Wise} and the DSS show that the star is rather isolated (Fig~\ref{Skyview}). This supports the very low contamination rate that was derived for the star within the CoRoT photometric
mask (see Sec~\ref{SysParam}).

  \begin{figure}
   \centering
    \includegraphics[height=6cm,width=9cm]{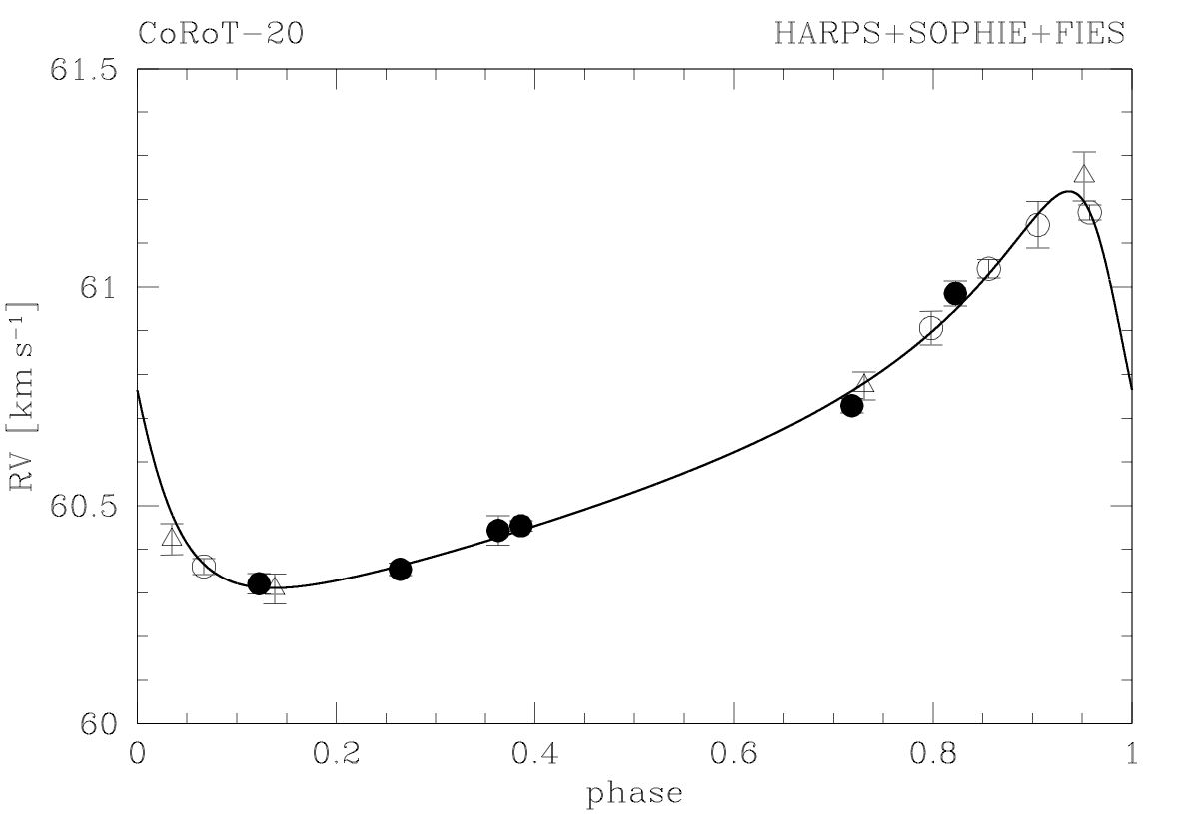}
   \caption{The phase-folded radial velocity measurements of \textit{CoRoT-20}. The various 
symbols correspond to the different spectrographs used for the follow-up campaign: 
\textit{HARPS} (black circle), \textit{SOPHIE} (open circle) and \textit{FIES} (open triangle). 
The best fit solution is over-plotted in full line}
 \label{RadVel}
 \end{figure}
\begin{figure}[]
\begin{center} 
\includegraphics[height=4cm,width=9cm]{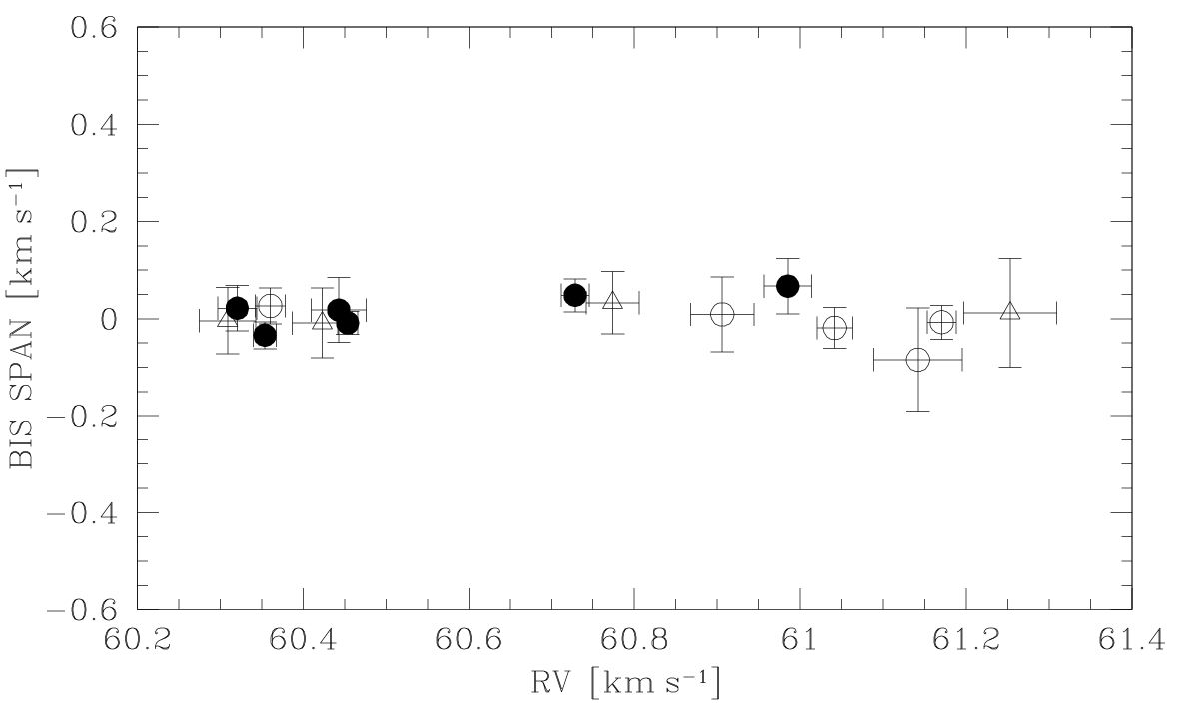}
\end{center}
\caption{Bisector span versus radial velocity of \textit{CoRoT-20} showing no correlation.}
\label{Bisector}
\end{figure}

Radial velocity (RV) observations started during the same season on December 9, 2010.  We used the \harps\ spectrograph \citep{Mayor03} mounted on the 3.6-m ESO telescope (Chile)  as part of the ESO large program 184.C-0639, the \sophie\ spectrograph  \citep{Perruchot08} on the 1.93-m telescope at the Observatoire de Haute Provence 
(France) and the \fies\ spectrograph on the Nordic Optical Telescope \citep{Frandsen99} based on the 2.56-m Nordic Optical telescope in La Palma (Spain) under observing program P42-216. 
\begin{table}[ht]
\begin{center}{
\caption{\label {TabRV}  Log of radial velocity observations}
\begin{tabular}{lllll}
\hline
\hline
Date  & HJD    & \vrad\  & $\sigma$\vrad  & Spectrograph \\
          &           & \kms    &  \kms                &    \\
\hline
2010-12-09 & 55540.70645 & 60.728 & 0.0169 & \harps\ \\
2010-12-14 & 55545.75533 & 60.353 & 0.0138 & \harps\ \\
2011-01-12 & 55574.60927 & 60.453 & 0.0119 & \harps\ \\
2011-01-16 & 55578.64264 & 60.985 & 0.0285 & \harps\ \\
2011-01-21 & 55583.64035 & 60.443 & 0.0334 & \harps\ \\
2011-01-28 & 55590.66159 & 60.320 & 0.0234 & \harps\ \\
\hline
2011-01-16 & 55578.41934 & 60.813 & 0.0385 & \sophie\ \\
2011-01-17 & 55579.41057 & 61.048 & 0.0532 & \sophie\ \\
2011-02-04 & 55597.44200 & 60.948 & 0.0213 & \sophie\ \\
2011-02-05 & 55598.38017 & 61.077 & 0.0175 & \sophie\ \\
2011-02-06 & 55599.39134 & 60.267 & 0.0183 & \sophie\ \\
\hline
2011-01-06    &   55568.55091   &  60.611  &   0.032 	  &   \fies\  \\
2011-01-08    &   55570.59973   &  61.090  &   0.060     &    \fies\   \\
2011-01-18    &   55580.60213   &  60.260  &   0.036     &    \fies\   \\
2011-01-19    &   55581.56506   &  60.146  &   0.034     &    \fies\   \\
\hline
\end{tabular}}
\end{center}
\end{table}

We used the same instrument set-up as for previous \corot\ candidates follow-up :  
high resolution mode for \harps\ and high efficiency mode for \sophie\ 
without acquisition of the simultaneous thorium-argon calibration, the second fiber being 
used to monitor the Moon background light \citep{Santerne11}.  For \textit{HARPS} and 
\sophie, the exposure time was set to 1 hour. We reduced data and computed RVs 
with the pipeline based on the weighted cross-correlation function (CCF) using a numerical 
G2 mask \citep{Baranne96, Pepe02}. 

\textit{FIES} observations were performed in high-resolution mode with the 1.3 arcsec fiber 
yielding a resolving power $R\approx67\,000$ and a spectral coverage from  
3600 to 7400\,\AA. Three consecutive exposures of 1200 sec were obtained 
for each observation. Long-exposed ThAr spectra were acquired right before 
and after each science spectra set, as described in \cite{Buchhave10}. 
Standard IRAF routines were used to reduce, combine, and wavelength calibrate the
nightly spectra. RV measurements were derived cross-correlating the
science spectra with the spectrum of the RV standard star HD\,50692 \citep{Udry99}, observed with the same instrument set-up as \Cttws. 

The 15 radial velocities of \Cttw\ are listed in Table~\ref{TabRV} and displayed in Fig~\ref{RadVel}. 
They present a clear variation, in phase with the \corot\ 
ephemeris and consistent with a companion in the planet-mass regime with an eccentric orbit. 
We nevertheless investigated the possibility of an unresolved eclipsing binary being the source 
of observed transits. To that purpose, we performed the line-bisector analysis of the CCFs 
(see Fig.~\ref{Bisector}) and also checked that there is no dependency of the RVs variations with the 
cross-correlation masks constructed for different spectral types \citep{Bouchy09}. 

The Keplerian fit of the RVs was performed simultaneously with the transit modeling (see Section 4). 
All the parameters of the fit are listed in Table~\ref{AllParams}.

\section{System parameters}
\label{SysParam}
 We calculated the flux contamination from nearby stars whose light might fall inside the \corot\ photometric mask using the same methodology as described in \cite{Borde10}. The method takes into account the photometric mask used to perform the on-board photometry and all the stars in the target neighborhood, including  faint background stars. We found the contamination being less than 0.6\% and we further neglected it. 

Three sections of the light curve, each centered on a transit, were locally normalized by fitting a third-degree polynomial. Each section was a 5-hours interval before the transit ingress and after its egress.
The detailed physical modeling of the system was performed by carrying out  the transit modeling and the Keplerian fit of the radial velocity measurements simultaneously. For the transit fit we used the formalism of \citet{Gimenez06, Gimenez09}. 
The fit implies twelve free parameters :  the orbital period $P$, the transit epoch $T_{\rm tr}$, the transit duration $T_{\rm 14}$, the ratio of the planet to stellar radii \Rp/\Rs, the inclination $i$ between the orbital plane and the plane of the sky,  the Lagrangian orbital elements $h=e~\sin{\omega}$ and $k=e~\cos{\omega}$, where $e$ is the eccentricity and $\omega$ the argument of the periastron, the radial-velocity semi-amplitude $K$, the systemic velocity $\gamma_{rel}$ and the two offsets between \sophie\ and \harps\ radial velocities and \sophie\ and \fies. For the transit modeling, we used a limb-darkening quadratic law \citep{2003A&A...401..657C,2004A&A...428.1001C}. The limb-darkening coefficients $u_{a}$ and $u_{b}$ were taken using the tabulated values for the \corot\ bandpass from \citet{Sing10} for the atmospheric  parameters \teff, \logg\  and metallicity derived for the central star (see sect.~\ref{StellarP}): $u_{a} = 0.4262 \pm 0.0168$ and $u_{b} = 0.2434 \pm 0.0108$). The two corresponding non-linear limb-darkening coefficients are $u_{+} = u_{a} + u_{b} = 0.6696 \pm 0.0201 $ and $u_{-} = u_{a} - u_{b} =  0.1828 \pm 0.0201$. We decided to keep these limb-darkening parameter values fixed in the transit fitting. 

\begin{figure}[h]
\begin{center} 
 \includegraphics[width=6cm,height=9cm,angle=90]{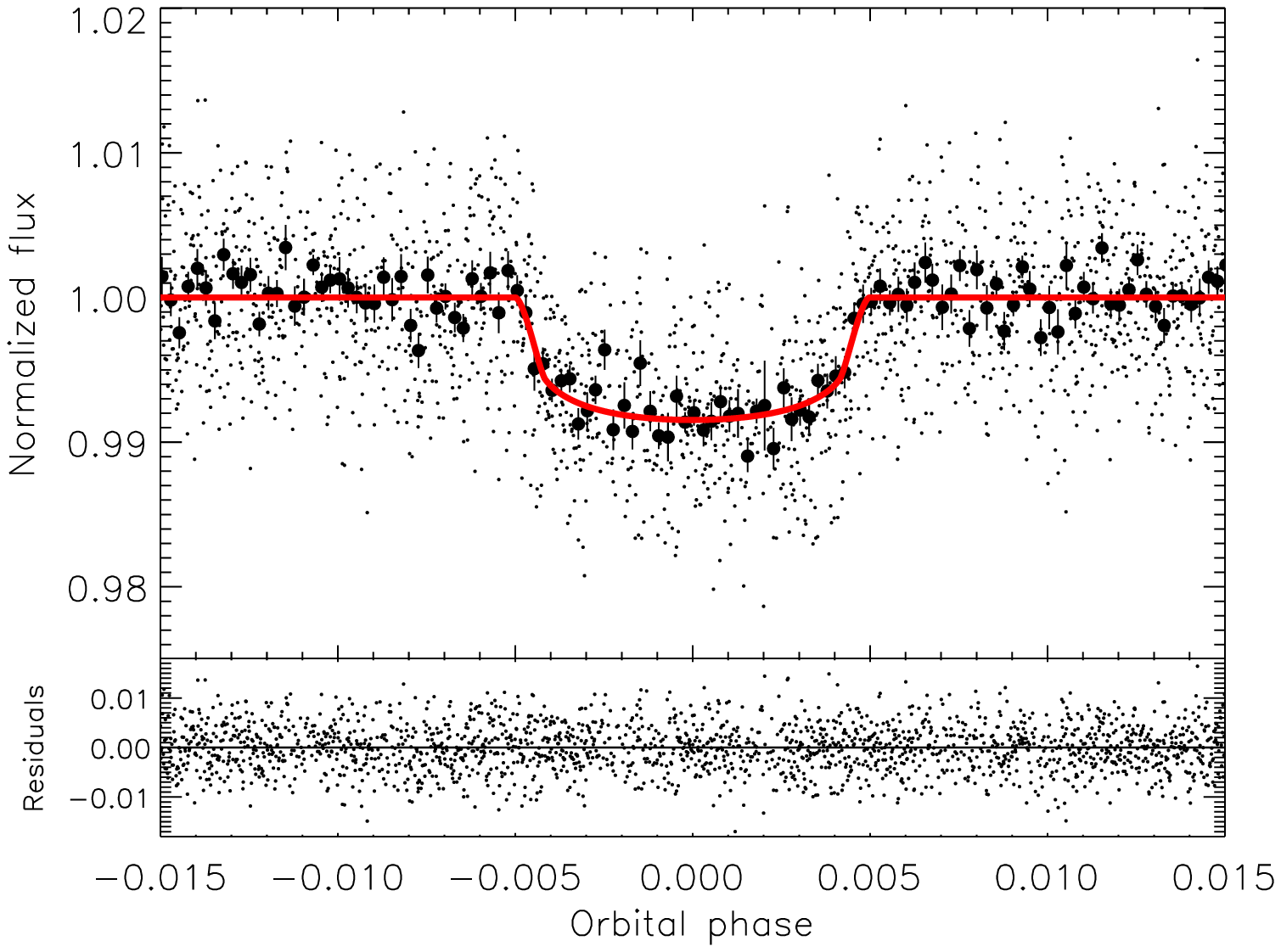}
\end{center}
\caption{The phase-folded transit in the phase space. The phase bins are 3.3~min and the error bar of each individual bin was calculated as the dispersion of the points inside the bin, divided by the square root of the number of points per bin. The best model is over plotted in full line.}
\label{TransitFit}
\end{figure}

The fit was performed using the algorithm AMOEBA \citep{Press92}. The initial values of the fitted parameters  were changed with a Monte-Carlo method to find the global minimum of the $\chi^{2}$. The associated 1-sigma errors were then estimated using a bootstrap procedure described in details in \cite{Bouchy11}. In such a procedure the limb-darkening parameters were allowed to vary within their error bars related to the atmospheric parameter uncertainties. The final values of the fitted parameters and the subsequently derived system parameters are given in Table~\ref{AllParams}. Fig.~\ref{TransitFit} displays the best fit compared to the observed folded transit.

\section{Stellar parameters}
\label{StellarP}

The spectroscopic analysis was done the usual way it is carried out for the \corot\ planets: a master spectrum was created from the co-addition of spectra collected for the radial velocity measurements of the companion. We chose the \harps\ spectra that offer the highest spectral resolution. We selected those that were not affected by the Moon reflected light at the time of the observations. Each order of the selected spectra was corrected by the blaze, set in the barycentric rest frame and rebinned to the same wavelength grid with a constant step of 0.01\AA. The spectra were then co-added order per order. Each order of the co-added spectrum was then carefully normalized and the overlapping orders were merged resulting in a single 1D spectrum. This master spectrum has a S/N of 176 per element of resolution at 5760\AA\ in the continuum. 
 \begin{figure}
 \begin{center} 
    \includegraphics[height=8cm,width=4cm,angle=90]{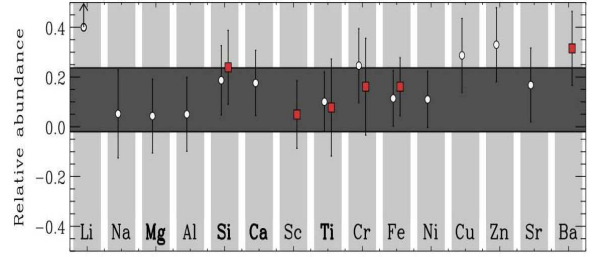}
   \caption{Abundances of the chemical elements measured with VWA in
the HARPS co-added spectrum of \Cttws. The abundances refer to the solar value. White circles correspond to neutral
lines, red boxes to singly ionized lines and the yellow area represents the mean metallicity within one sigma
error bar.}        
\end{center}      
\label{AbundCtw}
\end{figure}

 A prior estimate of the atmospheric parameters \teff, \logg, chemical composition and \vsini\  was performed by fitting  the spectrum with a library of synthetic spectra calculated using MARCS stellar atmosphere, including the \halpha\ Balmer line. The rotational broadening was estimated on a selection of isolated spectral lines fitted by synthetic spectra convolved with various rotational velocities. We found \vsini\ = 4.5 $\pm$ 1.0 \kms and \vmac\ = 3.5 $\pm$ 1.0 \kms.
The detailed analysis was then carried out using the Versatile Wavelength Analysis package (VWA) \citep{Bruntt04,BrunttDeleuil10}. A first set of weak and isolated lines of \ion{Fe}{i} and \ion{Fe}{ii} was fitted until the derived abundances of Fe minimized the correlation with the equivalent width and the excitation potentials. We found : \teff\ = 5880 $\pm$ 90 K, \logg\ = 4.05 $\pm$ 0.17 and \vmic\ = 1.10 $\pm$ 0.1 \kms\ which corresponds to a G2-type dwarf. Then the abundances of other elements for which we could find isolated spectral lines were derived (see Fig~\ref{AbundCtw}). We performed an independent estimate of the surface gravity from the pressure-sensitive lines: the \ion{Mg}{i}1b lines, the \ion{Na}{i} D doublet and  the \ion{Ca}{i} at 6122\AA\ and 6262\AA. We fitted the spectrum with the aforementioned  grid of synthetic spectra in regions centered on each of the spectral lines of interest. The inferred value of the surface gravity is \logg\ = 4.2$\pm 0.15$, a value in good agreement with the \logg\ derived with VWA obtained from the agreement between the \ion{Fe}{i} and \ion{Fe}{ii} abundances.  We thus adopted \logg = 4.2 for the surface gravity. 

The mean metallicity of the star is computed as the mean of metals with more than 10 lines in the spectrum, such as Si, Ca, Ti, Fe, Ni (Fig.~\ref{AbundCtw}). This yields a straight mean of ${\rm [M/H]} = 0.14\pm  0.05$. The error on [M/H] due to the uncertainty on \teff, \logg\ and microturbulence is 0.11 dex, which we must add quadratically to get ${\rm [M/H]} = 0.14\pm 0.12$ \citep{BrunttBedding10}.

\begin{figure}
   \centering
    \includegraphics[height=9cm,width=4cm,angle=90]{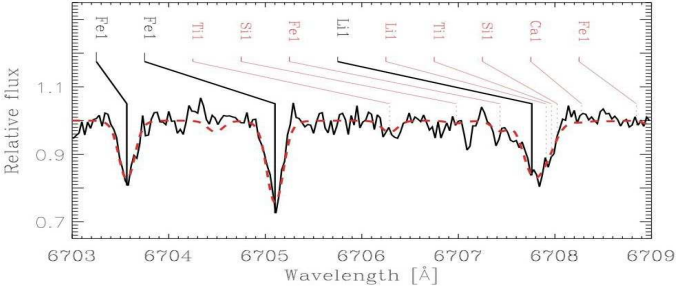}
   \caption{\Cttws\ spectrum in a spectral region around the \ion{Li}{i} lines at 6708\AA. }              
\label{LiLine}
    \end{figure}

We also checked for any indicators of age. We found no hint of stellar activity in the \ion{Ca}{ii} H and K lines. However, the \ion{Li}{i} line is clearly detected at 6708\AA\ (see Fig.~\ref{LiLine}). We measured an equivalent width of $W_{eq}$ = 44~m\AA\ and determined a lithium abundance of 2.97. Following \cite{Sestito05}, this leads to an age in the range 100~Myr up to 1~Gyr, depending on the star's initial rotation velocity.

The modeling of the star in the HR diagram was carried out in the (\teff,\mtier) plane taking the host-star's metallicity into account. It resulted in the final estimates of the star's fundamental parameters given in Table~\ref{AllParams}: \Ms\ = 1.14 $\pm$ 0.08 $M_\odot$, \Rs\ = 1.02 $\pm$ 0.05 \Rsun. The inferred surface gravity is \logg\ = 4.47 $\pm$ 0.11, in agreement within the errors with the spectroscopic result. The evolutionary status points to a young star likely in the last stages of the pre-MS phase. We found the most likely isochrone age being 100$_{-40}^{+800}$ Myr, a result in good agreement with the Li abundance.

We calculated the distance of the star. We used the parameters of the star we derived and its \mass\ magnitudes to estimate the reddening. We found a color excess $E(J-K)$ = 0.18~mag and the absorption  $A_{V} = 1.04 \pm 0.5 ~mag$ using the extinction law from \cite{Schlegel98}. This yields a distance of 1.23 $\pm$ 0.12 kpc, consistent with the strong interstellar absorption observed in the \ion{Na}{i} (D1) and (D2) lines.


\section{\Cttws\ system properties}
\label{SysProp}
Compared to the sample of known transiting planets, \Cttw\ is unusual in many respect.  With an orbital period of 9.24 
days it joins the group of transiting planets with periods outside the pile-up  at 3 days. It is the fifth planet discovered by \corot\ in this period domain which currently accounts for 25 planets (see http://exoplanet.eu/), 9 out of these belonging to multi-planet systems : Kepler-9 \citep{Holman10}, Kepler-10 \citep{Fressin11} and Kepler-11 \citep{Lissauer11}. However all these Kepler-planets have a mass which is less than $\sim$ 0.3 \MJ\ and could not be directly compared to the giant planet population. Excluding these planets in multiple systems, for the 17 remaining objects of the sample that do not have a detected companion, 8, that is 47\%, have a significant eccentric orbit with $e$ in the range 0.15 to 0.9. 

Planets with highly eccentric orbit appear to be found preferentially among the high-mass and/or long period planet population. With a mass of 4.13 \MJ\ which  places it at the border of the gap in mass between the regular hot-Jupiter population and the very massive planet one, \Cttw\ is consistent this trend. In the mass-period diagram they are clearly separated  from the lighter planets with circular orbits \citep{Pont11}. This dichotomy and in particular the lack of massive close-in planets at circular orbit suggest that tidal evolution should play an important role in the fate of the planet population. 

\subsection{Tidal evolution}

Following \cite{Levrard09} approach we checked the stability of \Cttw\ to tidal dissipation. The authors calculated the ratio between the total angular momentum of a given system $L_{tot}$ and the critical angular momentum $L_{crit}$ for some transiting systems. According to \cite{Hut80}, tidal equilibrium states exist when the total angular momentum is larger than this critical value $L_{crit}$. However, this equilibrium state could be stable or unstable, depending whether the orbital angular momentum $L_{orb}$ is more than three time the total spin angular momentum $L_{spin}$, or not. \cite{Levrard09} demonstrated that for none of the systems but HAT-P-2b the stable tidal equilibrium state, that corresponds to $L_{tot} / L_{crit} > 1$,  exists. Further the fate of these close-in planets is ultimately a collision with their host-star.  The study has been recently reexamined and extended to more than 60 transiting systems by \cite{Matsumura10} who achieved a similar conclusion, showing that the vast majority of these close-in planets will spiral-in to their host star and will be destroyed by tides. Using equations (1) and (2) given by \cite{Levrard09} that neglect any effect of a possible magnetized stellar wind, we found for \Cttw:
$$L_{tot} / L_{crit} = 1.057 {\rm \quad and \quad  } L_{spin\star} / L_{orb} = 0.0458$$ 
It shows that, within the current observational uncertainties, the planet has a tidal equilibrium state. It is worth noticing that our approach also assumes the stellar obliquity is small. The later is poorly constrained as the star's rotation period could not be derived from the light curve. We simply assumed that the rotation axis is perpendicular to the line of sight and derived the star's rotation period from the \vsini\ (Table~\ref{AllParams}), a regular method for transiting systems. This gives a rotational period of the star of 11.5 $\pm$ 3.1 days, that is of the same order than the planet's orbital period. In the case of \Cttw, $L_{spin\star} / L_{orb} < 1/3$ and most of the angular momentum of the system is in the orbit. According to \cite{Matsumura10}, \Cttw\ belongs to the very small subgroup of Darwin-stable systems that evolve toward a stable tidal equilibrium state 
where migration will stop. 

From the Roche-limit separation, the planet thus lies well beyond two times the Roche limit distance. Using  \citep{Faber05} :
$$a_R = (R_p/0.462) (M_\star / M_p)^{1/3}$$
we found that the Roche limit ${a_R}$ of the system is 0.0057 AU. This further supports the migration scenario over the scattering/Kozai-cycle scenario as proposed by \cite{Ford06}. 
\begin{figure}
\centering
\includegraphics[width=9cm,height=7cm]{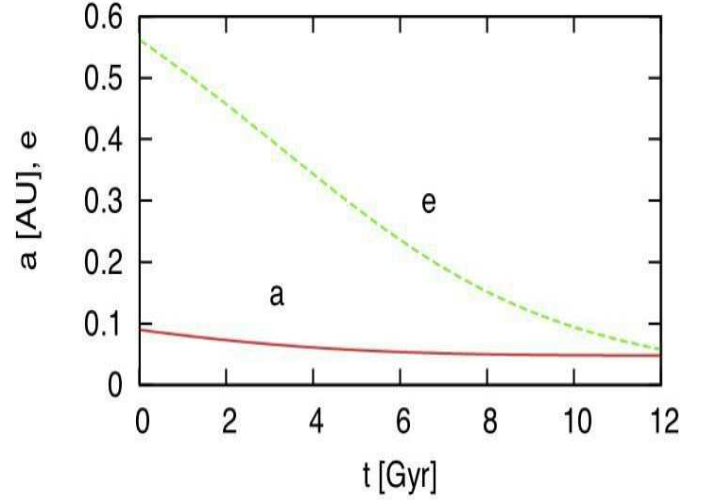}
\caption{Tidal evolution of the orbital semi-major axis and eccentricity. The figure is displayed on a time interval larger than the expected lifetime of the star to show the triple synchronization characteristic of a Darwin-stable system.}              
\label{TidalEvol}
\end{figure}
\begin{figure}
\centering
\includegraphics[width=9cm,height=7cm]{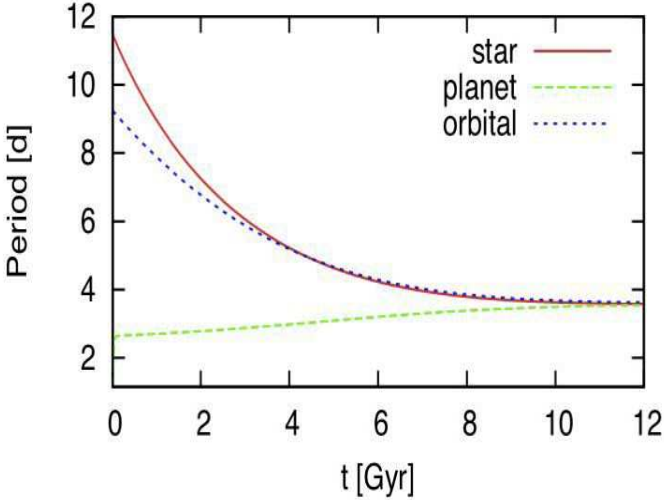}
\caption{Tidal evolution of the rotational and orbital periods.}              
\label{PeriodEvol}
\end{figure}

We performed a complete calculation of the tidal evolution of the system formed by the star and the planet assuming a linear tidal model \citep{Mignard79, Hut81}. The main difficulty here is to choose the values of the dissipation in the star and in the planet. For the main semi-diurnal tides of the star, we have adopted the value $Q^\prime_s=10^7$ as found for hot Jupiters \citep{Hansen10, Benitez-Llambay11}. Because of the close values of the orbital period and the rotation of the star, the components of the tides raised on the star by the planet related to the orbital eccentricity are also equally important, but the values of the current dissipation obtained with them are of the same order of magnitude. For the planet, we have derived one value on the basis of the actually determined $Q^\prime$ of Jupiter ($Q^\prime =1.36 \times 10^5$, see \cite{Lainey09}). We first note that standard linear tidal theories \citep[see][eqn 45]{Hut81} allow us to determine the current rotation period (stationary) of the planet independently of the dissipation. We obtain 2.64 $\pm$ 0.13 days. To transform $Q^\prime_{\rm Jup}$ into the planet's $Q^\prime_p$, we have to take into account: (i) $Q^\prime_p$ scales with the semi-diurnal tide period \citep[see][]{Ferraz-Mello08, Matsumura10}; (ii) $Q^\prime_p$ scales with $R_p^{-5}$ \citep[see][]{Eggleton98, Ogilvie04}. We thus obtain $Q^\prime_p=2.2 \times 10^6$. 
Fig.~\ref{TidalEvol} shows the variation of the semi-major axis and the evolution of the eccentricity. One can notice that with the adopted dissipation values, while the eccentricity tends toward zero, the circularization will not be achieved within the lifetime of the system.

Fig.~\ref{PeriodEvol} shows the evolution of the periods. The planet rotation is currently in a stationary super-synchronous state, that is the planet  rotation is faster than its orbital motion. Its period increases as the eccentricity decreases and almost synchronization is reached when the eccentricity becomes very small. The star rotation period is currently decreasing; it will equal the synchronous value at some time 4 Gyr from now. However, it will continue to decrease up to reach the triple synchronous stationary state. The triple synchronization, however, does not seem to be reached within the lifetime of the star.

It is worth underlining that the actual $Q^\prime$ values are not known and the values we used are only estimates founded on previous studies. Therefore the exact timescale of the tidal processes is uncertain. Furthermore, by extracting angular momentum from the system, stellar magnetic braking may prevent the planet from reaching a triple synchronous state and ultimately jeopardize its survival \citep{Bouchy11}. Indeed, simulations in which magnetic braking was active during the whole system lifetime, following the model proposed by \citep{Bouvier97} and using the same tide parameters as in the examples given above, show that the planet is falling below the Roche limit in about 6 Gyr. This result is critically dependent on the adopted parameters and further would required a detailed study that is well beyond the scope of the present paper.

We also investigate the consequences of the circularization of the planet orbit which is in the phase of fast circularization,  on the transits occurrence. Assuming there is no other close massive perturber in the system, then two effects are causing TTVs: the decrease of the orbital semi-major axis and the circularization of the orbit. Concerning the orbital semi-major axis, the time-scale of its variation, $\dot a$, is $-0.95 10^{-5}$ 1/Myr presently (see Fig~\ref{PeriodEvol}). As a consequence, a continuous period variation of $\dot P/P \approx -4 10^{-12}$ per cycle is expected. As well-known, this linear period variation will cause a parabolic $O-C$ curve, and in  100 years from now the $O-C$ value will be only $-25$ seconds. This is slightly over the $3\sigma$ observation limit by \corot\ \citep{Bean09, Csizmadia10}. Assuming that the transit timing precision can be forced down to $5$ sec in the future, this $O-C$ value will be reached in 45 years from now.

Turning to the evolution of the eccentricity during the circularization process, it has two consequences. First the occurrence
of the secondary eclipse will change. The displacement $D$ of the secondary from phase $0.5$ is given by \cite[eqn 1 and 2][e.g.]{Borkovits04}.
The previous results of the tidal evolution calculations indicated $\dot e = -4.5~10^{-5}$ 1/Myr and $\dot P = -1.5~10^{-3}$ days/Myr. Assuming a constant $\omega$, we have that $\dot D = -37.56~10^{-5}$ days/Myr or $\dot D = -9.53~10^{-12}$ days/cycle. This variation is of the same order as the previous one caused by the decreasing semi-major axis, so it would be observable within a century, too.

For the second effect, that is the circularization of the orbit, one can also consider the occurrence of a small precession of the orbit. This effect is hardly observable, but interesting on the theoretical side, since the transit occurs at the true anomaly
$v = 90^\circ - \omega$ where $\omega$ is the argument of periastrion.  
The later is also subject to variations because of theory of general relativity but also because
the tidal effects force the apsidal line to rotate. However, this variation has a different time-scale. We thus do not take this into account here, even if tidal forces also cause a small precession of the orbit showing that $\dot \omega$
is not zero. So if $e$ decreases due to circularization, and even
if $\omega$ is constant, then at the epoch of transit the eccentric anomaly will increase and hence the mean anomaly at transit will occur later. However, a first estimation shows that this effect may be negligible in a ten year timescale.

\subsection{Internal structure}
\Cttw\ is a massive hot-Jupiter with a mass of 4.24 \MJ\ a radius of 0.84 \MJ, and an inferred density $\rho = 8.87 \pm 1.1 \gcm$. A few giant planets are already reported with similar density or even higher :  \Ctfourt\ $\rho = 7.3 \pm 1.5 \gcm$ \citep{Tingley11}, WASP-18b $\rho = 8.8 \pm 0.9$ \citep{Hellier09,Southworth10} or HAT-P-20b $\rho = 13.78 \pm 1.5$ \gcm \citep{Bakos10} for example. While the mass of these planets spans a large range, from $\sim$ 4 up to more than 9\MJ, their radius is close to 1\RJ. Given \Cttw's large planetary mass, its small size is surprising. Among these high density giants planets, only HAT-P-20b has a comparable size, i.e. $0.867\pm 0.033$ \RJ. \Cttw, as HAT-P-20b, is thus expected to contain large amounts of heavy elements in its interior.

\begin{figure}
\centering
\includegraphics[width=9cm,height=7cm]{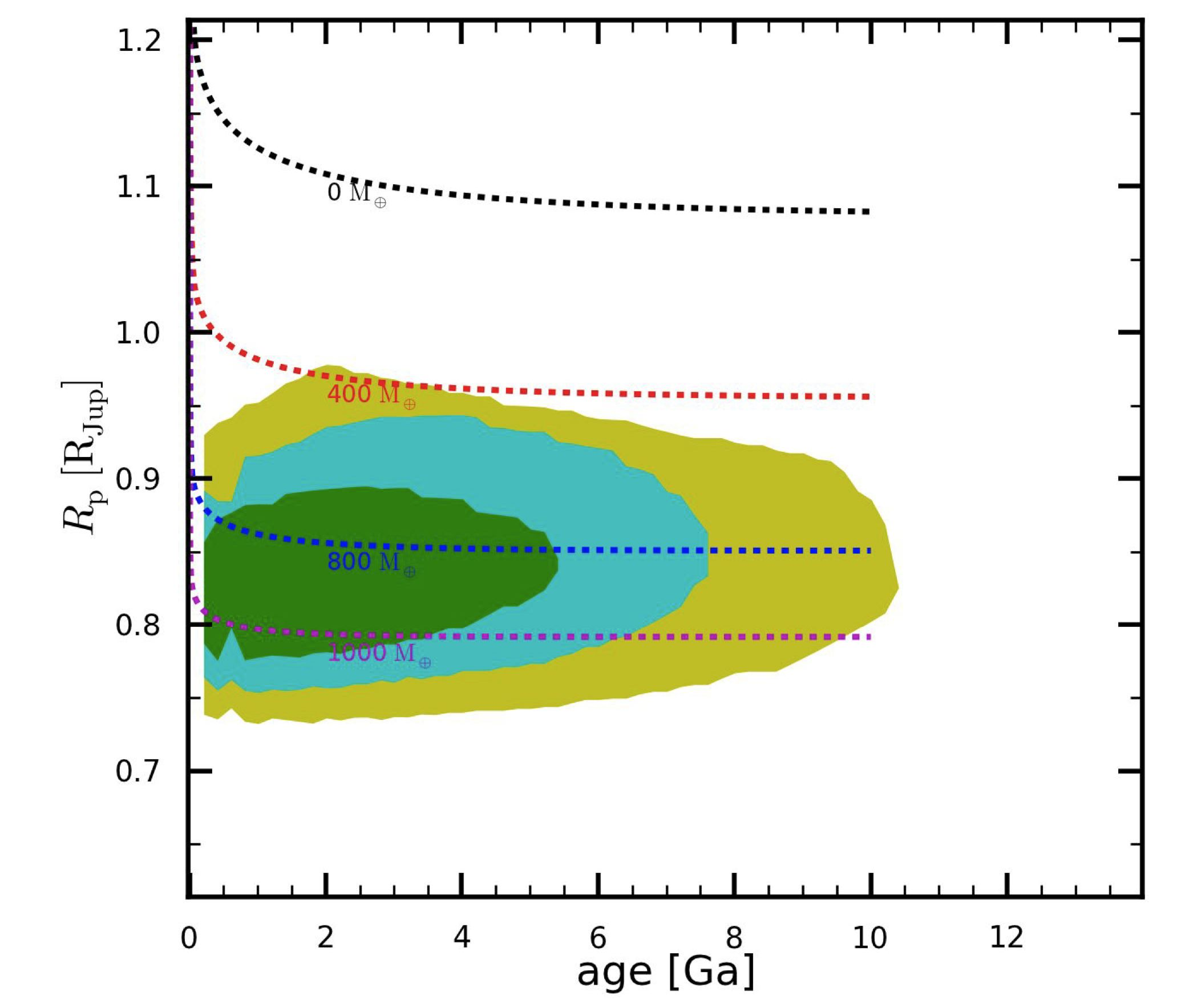}
\caption{Evolution of the size of \Cttw\ (in Jupiter units) as a function of age (in billion years), compared to 
constraints inferred from \corot\ photometry, spectroscopy, radial velocimetry  and stellar evolution models. Green, blue and yellow regions correspond to the  planetary radii and ages that result from stellar  evolution models matching 
the inferred $\rho_\star$ - \teff\ - \feh\ uncertainty ellipse within 1$\sigma$,  2$\sigma$ and 3$\sigma$, respectively. Planetary evolution models for a planet with a solar-composition envelope over a central dense core of variable mass (0, 400, 800, and 1000 \ME\ as labelled) are shown as  dashed lines. These models also assume that 1\% of the incoming stellar irradiation is dissipated deep into the interior of the planet.}
\label{Struct}
\end{figure}

To investigate the internal structure of \Cttw, we computed planetary evolution models with CEPAM \citep{Guillot95}, following the description in \cite{Guillot11}, and \cite{Havel11} for a planet of a total mass 4.24 \MJ. We derived a time-averaged equilibrium temperature of the planet to be $T_{\rm eq} = 1002 \pm 24$ K. The results for $T_{\rm eq}=1000\,$K are shown in Fig.~\ref{Struct} in terms of the planetary size as a function of the system age. The coloured regions (green, blue, yellow) indicate the constraints derived from the CESAM stellar evolution models \citep{Morel08} at 1, 2, and 3$\sigma$ level, respectively. For preferred ages between 100 Ma and 1 Ga, we find that \Cttw\ should contain between 680 and 1040 \ME\ of heavy-elements in its interior (i.e. between 50 and 77\% of the total planetary mass), at 1$\sigma$ level, about twice the amount needed for HAT-P-20b \cite[see][]{Leconte11OHP}. While this result is qualitatively in line with the observed correlation between star metallicity and heavy elements in the planet \citep[e.g.][and references therein]{Guillot06, Miller11}, the derived amounts are extremely surprising. They would imply that all the heavy elements of a putative gaseous protoplanetary disk of 0.1 to 0.15\,M$_\odot$ were filtered out to form \Cttw, and then that an extremely small fraction of hydrogen and helium in that disk was accreted by the planet. This is at odds with todays accretion formation models \cite[e.g.][]{Ida04, Mordasini09}. Disk instability with differentiation and partial tidal stripping  \citep{Boley10ApJ} is proposed as an alternative formation pathway. According to \cite{Boley11}, this could account for giant planets with massive core such as HAT-P-20b. In its current orbit, \Cttw\ doesn't enter the Roche limit and no tidal stripping is acting on yet but this scenario would deserve further investigations.

We investigated the possibility that changes in the atmospheric model would yield more "reasonable" values for the planetary enrichment. As can be  seen from a similar study in the brown dwarf regime \citep{Burrows11}, the consequences of modified atmospheric properties are limited for objects with  the mass of \Cttw\ (i.e. standard radii for objects of this mass range from  1.05 to 1.20$\,\rm R_{Jup}$). By artificially lowering the infrared  atmospheric opacity by a factor 1000 (not shown), we were able to decrease the $1\sigma$ upper limit to the core mass from $650$ to $390\,\rm M_\oplus$, a small change compared to huge and unphysical decrease in the opacity. 

On the other hand, one strong assumption in our study is that heavy 
elements are embedded into a central core. When relatively small amounts of 
heavy elements are considered, it is not very important whether they are 
considered as being part of a core or mixed in the envelope \citep[e.g.][]{Ikoma06}. However, as shown by \cite{Baraffe08}, when 0.5 \MJ\ of  ices are mixed in the envelope of a 1 \MJ\ planet, its radius is 
smaller by $\sim$ 0.1 \RJ\ than when one considers that these elements  are part of a central core. It is thus very likely that the mass of heavy elements required to explain the radius of \Cttw\ is high but significantly 
smaller than considered here. Estimates based on the \cite{Baraffe08} calculations indicate that if mixed in the envelope, a mass of heavy elements 2 to 3 times smaller than estimated in Fig.~\ref{Struct} would explain the observed planetary size. This would alleviate the problem of the formation of the  planet, although it would still require relatively extreme/unlikely scenarios.

\section{Summary}
In this article we presented the discovery of \Cttw. The object  belongs to the population of massive planets with orbital semi major axes below 0.1 AU, a domain of orbital periods where low and  high eccentricity  systems co-exist in a narrow range of orbital period.   We examined the tidal  stability of \Cttws\ and found that, within the observational uncertainties, 
it belongs the relatively small population of transiting planets that are  considered as "Darwin-stable", i.e. systems for which in the absence of  processes extracting angular momentum from the system (i.e. stellar winds),  the planet would never fall onto the central star which would instead be spun-up towards triple-synchronization (equality of the orbital, planetary and 
stellar spin periods). Other cases in the planetary  domain are \Ctthree, \Ctsix, WASP-7 and HD 80606 \citep{Matsumura10}. Measuring the stellar obliquity through the Rossiter-McLaughlin effect would provide additional constraints to better constrain its tidal evolution and further understand its orbital evolution. The expected semi-amplitude of the radial velocity anomaly of the Rossiter-McLaughlin effect is estimated to be 22 $\pm$ 5~\ms, unfortunately quite difficult to detect with the present spectrographs used for the keplerian orbit determination. Another point would be to assess the presence of any additional companions in the system by long-term radial velocity monitoring as, according to \cite{Matsumura10}, the formation path of close-in planet should be different for single-planet system and multi-planet ones. \Cttws\ system appears thus an interesting bench test case to study the tidal orbital and rotational evolution of the close-in population.

The second interesting peculiarity that distinguishes \Cttw\ from the regular giant planet population is its small observed radius.  According to planetary evolution models, the interior of this compact planet should contain a very high amount of heavy elements, with a central dense core whose mass would be in the range between 680 and 1040 \ME. Although mixing heavy elements in the envelopes rather than confining them to a central core can lead to substantially smaller values (by a factor estimated to be $\sim 2-3$), the origin of such a huge amount of heavy elements is difficult to explain within the framework provided by the current planetary formation models. With HAT-P-20b \citep{Bakos10}, it is the second example of extremely metal rich interior that challenges planetary interior models. However, the two planetary systems differ in many aspects: HAT-P-20 mass is nearly twice that of \Cttw; it orbits on a nearly circular orbit a K3 metal rich star, while \Cttws\ is a solar-type, slightly metal enriched star. In addition, HAT-P-20 seems to be physically associated to another stellar companion, while up to now, \Cttws\ has none, detected or suspected.  It thus appears difficult from this two exceptions to derive any trend that would provide hints on the origin these challenging and intriguing bodies.  

\vfill\eject
\begin{acknowledgements}
The French team thanks the CNES for its continuous support to the \corot/Exoplanet program.
The authors wish to thank the staff at ESO La Silla Observatory for their support and for their contribution to the
success of the HARPS project and operation. The team at the IAC acknowledges support by grants
ESP2007-65480-C02-02 and AYA2010-20982-C02-02 of the Spanish Ministry
of Science and Innovation (MICINN). The CoRoT/Exoplanet catalogue
(Exodat) was made possible by observations collected for years at the
Isaac Newton Telescope (INT), operated on the island of La Palma
by the Isaac Newton group in the Spanish Observatorio del Roque
de Los Muchachos of the Instituto de Astrophysica de Canarias. The
German CoRoT team (TLS and University of Cologne) acknowledges
DLR grants 50OW0204, 50OW0603, and 50QP0701. The
Swiss team acknowledges the ESA PRODEX program and the Swiss
National Science Foundation for their continuous support on CoRoT
ground follow-up. A. S. Bonomo acknowledges CNRS/CNES grant
07/0879-Corot. S. Aigrain acknowledges STFC grant ST/G002266.
M. Gillon acknowledges support from the Belgian Science Policy
Office in the form of a Return Grant.
\end{acknowledgements}

\bibliographystyle{aa.bst}
\bibliography{corot20}
\clearpage 

\begin{table}[ht]
\caption{ \label{AllParams} Planet and star parameters.}            
\begin{minipage}[!]{17.0cm}  
\renewcommand{\footnoterule}{}   
\begin{tabular}{ll}        
\hline\hline                 
\multicolumn{2}{l}{\emph{Ephemeris}} \\
\hline
Planet orbital period $P$ [days] & 9.24285 $\pm$ 0.00030 \\
Primary transit epoch $T_{tr}$ [BJD-2400000] &  55 266.0001 $\pm$ 0.0014 \\
Primary transit duration $T_{14}$ [d] & 0.0927  $\pm$ 0.0019 \\
Secondary transit epoch $T_{s}$ [BJD-2400000] & 55 272.46 $\pm$ 0.13\\
\multicolumn{2}{l}{\emph{System parameters}} \\
\hline    
Periastron epoch $T_{peri}$ [BJD-2400000]   &      55265.79074 \\
 $e {\rm \sin} \omega$  & 0.468 $\pm$ 0.017   \\
 $e {\rm \cos} \omega $  & 0.312 $\pm$ 0.022 \\
Orbital eccentricity $e$  &  0.562 $\pm$ 0.013\\
Argument of periastron $\omega$ [deg] & 56.3$_{-2.3}^{+2.4}$ \\ 
Radial velocity semi-amplitude $K$ [\ms] & 454 $\pm$ 9\\
Systemic velocity   $\gamma_{rel}$ [\kms] & 60.623 $\pm$ 0.006\\
\harps-\sophie\ offset velocity $V_{r1}$ [\ms]   &    93 $\pm$ 11  \\
\sophie-\fies\   offset velocity  $V_{r2}$ [\ms]   &     163 $\pm$ 20 \\ 
O-C residuals [\ms] &  27 \\
& \\
Radius ratio $R_{p}/R_{\star}$ & 0.0842 $\pm$ 0.0017 \\
Impact parameter\tablefootmark{a} $b$ & 0.26 $\pm$ 0.08 \\
Scaled semi-major axis $a/R_{\star}$~$^b$ & 18.95$_{-0.73}^{+0.63}$ \\
\mtier\ [solar units] & 1.022$_{-0.039}^{+0.034}$\\
Stellar density $\rho_{\star}$ [$g\;cm^{-3}$] &1.51$_{-0.17}^{+0.15}$ \\
Inclination $i$ [deg] & 88.21$\pm$ 0.53  \\
\multicolumn{2}{l}{\emph{Spectroscopic parameters }} \\
\hline
Effective temperature $T_{eff}$[K] & 5880  $\pm$  90\\
Surface gravity log\,$g$ [dex]& 4.20 $\pm$ 0.15 \\
Metallicity $[\rm{Fe/H}]$ [dex]& 0.14 $\pm$ 0.12\\
Stellar rotational velocity {\vsini} [\kms] & 4.5 $\pm$ 1.0 \\
Spectral type & G2 V \\
\multicolumn{2}{l}{\emph{Stellar and planetary physical parameters from combined analysis}} \\
\hline
Star mass [\Msun] &  1.14 $\pm$ 0.08 \\
Star radius [\Rsun] &  1.02 $\pm$ 0.05 \\
Distance of the system [kpc] & 1.23 $\pm$120  \\
Age of the star $t$ [Myr] & 100$_{-40}^{+800}$ \\
Orbital semi-major axis $a$ [AU] & 0.0902 $\pm$ 0.0021 \\
Orbital distance at periastron $a_{per}$ [AU] & 0.0392 $\pm$ 0.0017 \\
Orbital distance at apastron $a_{apo}$ [AU] &   0.1409 $\pm$0.0037 \\
Planet mass $M_{p}$ [M$_J$ ]$^c$ &  4.24 $\pm$ 0.23 \\
Planet radius $R_{p}$[R$_J$]$^c$  &  0.84 $\pm$ 0.04 \\
Planet density $\rho_{p}$ [$g\;cm^{-3}$] &  8.87 $\pm$ 1.10 \\
Equilibrium temperature $^d$  $T_{eq}$ [K] & 1002 $\pm$ 24 \\
Equilibrium temperature at periastron$^d$  $T^{per}_{eq}$ [K] & 1444$_{-46}^{+53}$\\
Equilibrium temperature at apastron$^d$ $T^{apa}_{eq}$ [K] & 764 $\pm$18 \\
\hline       
\vspace{-0.5cm}
\footnotetext[1]{$a/R_{\star}=\frac{1+e \cdot \cos{\nu_{1}}}{1-e^{2}} \cdot \frac{1+k}{\sqrt{1-\cos^{2}({\nu_{1}+\omega-\frac{\pi}{2}})\cdot \sin^{2}{i}}}$, where $\nu_{1}$ is the true anomaly measured from the periastron passage at the first contact \citep[see][]{Gimenez09}.}
\footnotetext[2]{$b=\frac{a \cdot \cos{i}}{R_{*}} \cdot \frac{1-e^{2}}{1+e \cdot \sin{\omega}}$}
\footnotetext[3]{Radius and mass of Jupiter taken as 71492 km and 1.8986$\times$10$^{30}$ g, respectively.}
\footnotetext[4]{zero albedo equilibrium temperature for an isotropic planetary emission.}
\end{tabular}
\end{minipage}
\end{table}

\end{document}